\documentclass[aps,preprint,superscriptaddress,showpacs]{revtex4}

\usepackage{amsmath,bm}
\usepackage{graphicx}

\begin{document}

\title{Reduction of activation energy barrier of Stone-Wales transformation in endohedral metallofullerenes}
\author{Woon Ih Choi}
\affiliation{School of Physics, Seoul National University, Seoul 151-747, Korea}
\author{Gunn Kim}
\affiliation{BK21 Physics Research Division, Seoul National University, Seoul 151-747, Korea}
\affiliation{Department of Physics, North Carolina State University, NC 27695-7518, USA}
\author{Seungwu Han}
\affiliation{Department of Physics, Ewha Womans University, Seoul 120-750, Korea}
\author{Jisoon Ihm}
\email[corresponding author. Electronic address:\ ]{jihm@snu.ac.kr}
\affiliation{School of Physics, Seoul National University, Seoul 151-747, Korea}
\date{\today }

\begin{abstract}

We examine effects of encapsulated metal atoms inside a C$_{60}$ molecule on the activation energy barrier to the 
Stone-Wales transformation using {\it ab initio} calculations.
The encapsulated metal atoms we study are K, Ca and La which nominally 
donate one, two and three electrons to the C$_{60}$ cage, respectively.
We find that isomerization of the endohedral metallofullerene via the Stone-Wales transformation can occur more easily than 
that of the empty fullerene owing to the charge transfer. 
When K, Ca and La atoms are encapsulated inside the fullerene, the activation energy barriers are lowered 
by 0.30, 0.55 and 0.80 eV, respectively compared with that of empty C$_{60}$ (7.16 eV). 
The lower activation energy barrier of the Stone-Wales transformation implies
the higher probability of isomerization and coalescence of metallofullerenes, which require a series of Stone-Wales transformations.

\end{abstract}

\maketitle

Since its discovery,\cite{Kroto1} the fullerene has been extensively studied both theoretically and experimentally
because of its potential applications in the fields of nanomaterials and biomedical science.
Although the fullerene molecule usually has the highest icosahedral symmetry allowed by sixty constituent
carbon atoms, transformations to other isomers of lower symmetry are still possible. 
One of the  most plausible processes for isomerization of the fullerene is the so-called Stone-Wales (SW)
 or ``pyracylene" rearrangement, which is the 90$^{\circ}$ rotation of two carbon atoms 
about the midpoint of the bond.\cite{SWT}
The SW transformation is also used to describe the structural changes 
of $sp^2$-bonded carbon nanosystems.\cite{Diederich}
For example, it has been proposed that the fusion process of fullerenes or carbon nanotubes may be explained in terms
of a sequence of this rearrangement.\cite{Hansw,Zhao,mnyoon}
A problem in this proposal is that the activation energy barrier of $\sim$ 7 eV,\cite{Bettinger} on the basis
of cutting two $\sigma$ bonds, is somewhat too high to overcome at experimental temperature for the merging process 
(1000$-$1500$^{\circ}$C).\cite{peapodfusion1, peapodfusion2} 
Nevertheless the transformation is useful to explain the production of all isomers of fullerenes and carbon nanotubes. 
Eggen {\it et al}. found that the activation barrier for this rearrangement was substantially reduced 
in the presence of a loosely bound carbon atom located
preferentially in the region of paired pentagons.\cite{Eggen} Therefore pentagon rearrangements which are necessary steps 
in the growth of fullerenes may considerably result from autocatalysis by carbon.
Osawa {\it et al}. reported the catalytic effect of various kinds of elements or ions in the bowl-shaped C$_{34}$H$_{12}$.\cite{osawa}
However, relatively few studies have been devoted to the SW transformation of the endohedral metallofullerene. 
In this paper, we investigate effects of encapsulated metal atoms on the activation energy barrier
in the SW transformation of the M@C$_{60}$ molecule, where M = K, Ca or La.
We shall show below that isomerization of the endohedral metallofullerene via the Stone-Wales transformation 
can occur more easily than that of the bare fullerene owing to the charge transfer from the encapsulated metal atoms
to the fullerene.
Experimentally, metallofullerenes are much easier to fuse than empty
fullerenes in peapods.\cite{okazaki,nori}

In the present study, we perform {\it ab initio} pseudopotential 
calculations using the plane wave basis set\cite{Ihm} 
within the local density approximation (LDA) for the exchange-correlation effect. 
The ultrasoft pseudopotential\cite{vanderbilt} is adopted with the cutoff energy of 30 Ry. 
The supercell size is 29$\times$29$\times$29 $a^3_{B}$,
where $a_{B}$ is the Bohr radius, which can keep the undesirable interfullerene
interaction sufficiently small.
For La, we treat all 11 valence electrons (5$s^2$5$p^6$6$s^2$5$d^1$) explicitly since the 5$s$ and 5$p$ orbitals
cannot be considered to be frozen as core electrons.\cite{Laasonen}
The scalar relativistic effects are taken into account because the effects are important in heavy elements such as La.
 We start the optimization 
of the La position on a three-fold axis 1.5 \AA~away from the center 
of the cage and obtain the stable position of the La atom in the C$_{60}$ cage.  
For K@C$_{60}$ and Ca@C$_{60}$, the starting positions for optimization are at the center\cite{yyoon} and 
on a five-fold axis 1.5 \AA~away from the center of the cage,\cite{LWang} respectively.
The atomic positions are relaxed until the normal component of the atomic force becomes less than 0.02 eV/\AA.
We use the nudged elastic band (NEB) method\cite{neb} to calculate the activation energy barrier 
in the SW transformation of the fullerene.
The NEB method is efficient in finding the reaction pathway with minimum energy when both the initial and final states are known. 
Regarding SW transformation of metallofullerene, we rotate the C-C dimer which is the closest to the metal atom (Ca and La).
Thirteen replicas are chosen including the initial and final ones to construct the elastic band.
We also calculate the climbing image of the highest energy replica to pick up the exact transition state 
which is regarded as a saddle point in the potential energy surface.


In the relaxation of the metal atom in C$_{60}$, the K atom prefers to be at the center of C$_{60}$.
On the other hand, the stable position of the La atom is off-center and the distance between the La atom and the nearest C atom is $\sim$ 2.5~\AA.
The Ca atom shows the same tendency as La.
Upon encapsulation of the La atom, the three-fold degeneracy of the t$_{1u}$ states is split into
2+1 since the off-centered La atom breaks the icosahedral symmetry of an isolated C$_{60}$.\cite{Hansw2}
Almost three electrons are transferred from the La atom to the doubly degenerate states (four-fold degeneracy including spins) of C$_{60}$.\cite{Lu,Hansw2} The two-fold degeneracy is split again by the charging of odd number of electrons.
It is found that a nearly symmetric pathway which has a C$_2$ symmetry appears 
at the saddle point (transition state) and the activation energy profile [shown in Figs. 1(a) and (b)].
Comparing C$_{60}$ and La@C$_{60}$, the rotating carbon dimer has a bond length of 1.23 (1.27) \AA, 
with 1.40 (1.39) \AA~bonds to each of the two nearest
carbon atoms for C$_{60}$ (La@C$_{60}$). This result is in agreement with the previous studies.\cite{Bettinger,Wales}

 We identify the atomic geometry of the transition state (TS)
in SW transformations of C$_{60}$ and La@C$_{60}$ as depicted in Figs. 3(a) and (b).
Table I lists the activation barriers ($E_a$) of the SW transformations 
in various $sp^{2}$-bonded carbon systems. 
$E_a$ in the SW transformation of the fullerene is lower than 
in the graphene layer (9.2 eV, the highest among the systems we study). Lowering of the barrier in fullerenes is due to 
the curvature effect associated with the deviation from the $sp^2$ bonding.\cite{mnyoon}
When K, Ca and La atoms are encapsulated inside the fullerene, the activation energy barriers are lowered
by 0.30, 0.55 and 0.80 eV, respectively compared with that of bare C$_{60}$.
Reduction of the activation energy is smaller than in the case of 
the autocatalysis and other covalent materials.\cite{Eggen,osawa} 
To examine the major effect of the encapsulated metal atom on the lowering of the activation barrier,
we perform calculations for fullerenes charged with one or two electrons. Note that we do not show the data for C$^{3-}_{60}$. The third donated electron is not bound to the fullerene. 
When a fullerene is charged, the activation barrier becomes lower than that of the neutral cage.
The barrier reduction is approximately proportional to the number of donated electrons.
$E_a$ in the SW transformation of C$^{-}_{60}$ (C$^{2-}_{60}$) is almost the same 
as that of K@C$_{60}$ (Ca@C$_{60}$).
Therefore, the main role of the incorporated metal atom is the electron donation. 

We find the reaction energy ($E_{r}=E_{product}-E_{reactant}$) shows the same trend as $E_a$ in Table I.
The C$_{2v}$-symmetry C$_{60}$ with two pairs of fused pentagons (obtained by the SW transformation) is 1.59 eV
higher in energy than the I$_h$-symmetry C$_{60}$. 
This value is in agreement with other group's results in the literature.\cite{Eggen2}
By contrast, for La@C$_{60}$, $E_{r}$ is 0.3 eV, which is much smaller than that of C$_{60}$.
It represents that an isomer of the metallofullerene which has adjacent pentagons can exist with higher probability 
than bare C$_{60}$. 
In the case of K@C$_{60}$, E$_{r}$ has an intermediate value. 
The K atom at the center of the cage does not move during the process of the SW transformation.
By contrast, the La atom in the fullerene moves by 1.0~\AA~ toward the paired pentagons in the process of dimer rotation. 
The displacement of the La atom induces a large elongation 
of the C$_{60}$ cage compared with the empty C$_{60}$ molecule.
The geometry of an empty fullerene is governed by the isolated-pentagon rule (IPR),\cite{Kroto2,Schmalz}
which says that the most stable structure is that in which all pentagons are surrounded by five hexagons.
The IPR-violating cages such as Sc$_2$@C$_{66}$ and Sc$_3$N@C$_{68}$, however, are stabilized by the electron transfer
between the encapsulated metal atoms and the carbon cage, which significantly decreases
the strain energy caused by paired pentagons and thus stabilizes the fullerene.\cite{CRWang,Stevenson}
Owing to the 108$^\circ$ bond angles of the pentagon,
the apex atoms in the fused (paired) pentagon regions of the empty fullerene
have a hybridization very close to $sp^3$ and exhibit a dangling-bond-like character 
which cannot be hybridized with any other electronic state.\cite{Kaxiras}
For metallofullerenes, on the other hand, the electron donation 
from the metal atom weakens the dangling-bond-like character.

To clarify the role of the encapsulated metal atom, we have investigated the electronic structures 
of C$_{60}$ and La@C$_{60}$ in detail.  
The projected density of states (PDOS) to the 6\textit{s} and 5\textit{d} orbitals of La for each reactant, transition state 
and product are drawn (not shown).  
It is observed that the energy levels of the La atom lie mostly above the Fermi level. 
It means that almost three electrons move from La to the fullerene in La@C$_{60}$.
In Fig. 2(a), the highest occupied molecular orbital (HOMO) at $-$0.2 eV corresponds to the dangling-bond-originated state 
as illustrated in Fig. 2(b). 
This state is above the HOMO of C$_{60}$ (i.e., the five-fold states originating from the
HOMO of empty C$_{60}$ which are split-off now). Energy level which has the dangling bond character in La@C$_{60}$ 
is at $-$0.7 eV (HOMO$-$2) as shown in Fig. 2(c) by arrow and the wave function in Fig. 2(d). 
Here we can see the three electrons coming from the La atom occupy the upper one and half level 
which is the lowest unoccupied molecular orbital (LUMO) and LUMO+1 state in the transition state of C$_{60}$. 
The overall shape of two wave functions [Figs. 3(b) and (d)] 
is the same with small change originating from the perturbation of the La atom.
Figures 3(c) and (d) show the PDOS into 2\textit{p}-orbitals of two carbon atoms 
[labeled A, B and A$'$, B$'$ in Figs. 3(a) and (b)] which have the dangling bonds in the transition state.
In the case of empty C$_{60}$, the PDOS of A and B are exactly the same. 
For La@C$_{60}$, on the other hand, the charge density
of nonbonding states at site A$'$ (between $-1$ and 0 eV) is not identical to the density at site B$'$.
This difference arises from the hybridization between the La atom and the C atom at site B.   
In the TS configuration of empty C$_{60}$, the HOMO and LUMO are both 
localized states within the hemisphere containing two dangling atoms (A and B).
By contrast, the TS geometry of La@C$_{60}$ has the HOMO and LUMO delocalized
on the whole cage.  
Consequently, the electron donation and hybridization between the fullerene and metal atom inside reduce
the activation energy barrier of the SW transformation in the endohedral metallofullerene.


We have studied the effect of encapsulated metal atoms inside a C$_{60}$ molecule 
on the activation energy barrier of the SW transformation 
using the {\it ab initio} density functional calculations.
The metal atoms in our study are K, Ca and La, which donate 
one, two and three electrons to the C$_{60}$ cage, respectively.
We have found that isomerization of the endohedral metallofullerene by the SW transformation can occur 
more easily than that of the empty fullerene owing to the charge transfer.
The reduction of the activation energy barrier by the electron transfer may also explain
the fact that metallofullerenes are much easier to coalesce than empty fullerenes inside nanotubes.

We thank Professor H. Shinohara for valuable discussions.
This work is supported by the CNNC of Sungkyunkwan University, 
the BK21 project of KRF, the Samsung SDI-SNU Display
Innovation, and the MOST through the NSTP (grant No. M1-0213-04-001).
The computations are performed at Supercomputing Center of KISTI 
through the Supercomputing Application Support Program using the PWSCF code.\cite{pwscf}

\newpage
\begin{center}
\LARGE{[Table Caption]}
\end{center}

Table I. Activation barrier ($E_a$) and reaction energy ($E_r$) of the SW transformation
in various $sp^{2}$-bonded carbon systems. All energy values are in eV.

\newpage
\begin{center}
\LARGE{[Figure Captions]}
\end{center}

Figure 1: Schematic energy diagram and the atomic configuration along the reaction coordinate in the Stone-Wales
transformation of empty C$_{60}$.
(a) Energy profile along the reaction coordinate. The dots represent the energies of the replicas in the SW transformation
in the NEB method. (b) Atomic rearrangement from the I$_{h}$ isomer 
to the C$_{2v}$ isomer of the empty C$_{60}$ molecule.

Figure 2: (color online) Density of states (DOS) for C$_{60}$ and La@C$_{60}$ in the transition state [(a) and (c) respectively] and the
isodensity surface plots of particular wave functions for C$_{60}$ and La@C$_{60}$ [(b) and (d)].
The energy levels of the wave functions in (b) and (d) are indicated with an arrow in (a) and (c), respectively.
The values for the red and blue isodensity surfaces are $\pm0.0014~e$/\AA$^3$.

Figure 3: (color online) Atomic structure of the transition state configuration for (a) C$_{60}$ and (b) La@C$_{60}$, and PDOS of two
atoms [labeled A, B and A$'$, B$'$ in (a) and (b)]
which have the remnant dangling bonds in the transition state for (c) C$_{60}$ and (d) La@C$_{60}$.
The defect states due to the dangling bonds and distorted bonding configurations are indicated by arrows.
The red solid and blue doted lines are the PDOS decomposed into 2{\it p}-orbitals of A, A$'$ and B, B$'$ atoms, respectively.

\newpage
\begin{table}

\begin{ruledtabular}
\begin{tabular}{ccc}
System & $E_a$ &   $E_r$ \\
\hline
    C$_{60}$                 &   7.16   &    1.59    \\
  K@C$_{60}$                 &   6.85   &    1.07    \\
 Ca@C$_{60}$                 &   6.61   &    0.59    \\
 La@C$_{60}$                 &   6.36   &    0.30    \\
 C$^{-}_{60}$               &   6.87   &    1.12    \\
 C$^{2-}_{60}$               &   6.59   &    0.67    \\
 C$_{42}$H$_{16}$ (for graphene)   &   9.20   &    3.07    \\
\end{tabular}
\end{ruledtabular}
\end{table}

\begin{center}
\LARGE{Table I}

\LARGE{W. I. Choi, G. Kim, S. Han and J. Ihm}
\end{center}

\end{document}